\documentclass{aa}

\usepackage{natbib}
\usepackage{graphicx}
\usepackage{txfonts}
\usepackage{epsfig}
 \bibpunct{(}{)}{,}{}{}{} 

\begin{document}
\title{The size distribution of magnetic bright points derived from Hinode/SOT observations}
\titlerunning{The size distribution of magnetic bright points}

\author{D. Utz
\inst 1
\and A. Hanslmeier
\inst 1
\and C. M{\"o}stl
\inst {1, 2}
\and R. Muller
\inst 3
\and A. Veronig
\inst 1
\and H. Muthsam
\inst 4}

\institute{IGAM/Institute of Physics, University of Graz, Universit{\"a}tsplatz 5, 8010 Graz, Austria
\and Space Research Institute, Austrian Academy of Sciences, Schmiedlstra\ss e 6, 8042 Graz, Austria
\and Laboratoire d\'{}Astrophysique de Toulouse et Tarbes, UMR5572, CNRS et Universit{\'e} Paul Sabatier Toulouse 3, 57 avenue d\'{}Azereix, 65000 Tarbes, France
\and Institute of Mathematics, Nordbergstra\ss e 15, 1090 Wien, University of Vienna, Austria}

\date{Received 28 August 2008 /
Accepted 14 January 2009}

\abstract{Magnetic Bright Points (MBPs) are small-scale magnetic
features in the solar photosphere. They may be a possible source of coronal heating by rapid footpoint motions that cause magnetohydrodynamical waves. The number and size distribution are of vital importance in estimating the small scale-magnetic-field energy.} {
The size distribution of MBPs is
derived for G-band images acquired by the Hinode/SOT instrument.} {For identification purposes, a new automated
segmentation and identification algorithm was developed.} {For a
sampling of 0.108 arcsec/pixel, we derived a mean diameter of
$(218 \pm 48)$~km for the MBPs. For the full resolved data set with
a sampling of 0.054 arcsec/pixel, the size distribution
shifted to a mean diameter of $(166 \pm 31)$~km. The determined
diameters are consistent with earlier published values. The shift is most probably due to the different spatial sampling.} {We
conclude that the smallest magnetic elements in the solar
photosphere cannot yet be resolved by G-band observations. The influence of discretisation effects (sampling) has also not yet been investigated sufficiently.}


\maketitle

\section{Introduction}
Magnetic Bright Points (MBPs) are small-scale magnetic features in the solar photosphere that are important for understanding solar magnetism. In addition, these features play a crucial role in solar physics in two ways. First of all, they can generate MHD waves that can contribute to the coronal heating problem, and secondly they can store magnetic energy.

To complete more robust statistical analyses and derive the parameters of
MBPs, an automated algorithm must be developed to identify the interesting features. Manually operated \citep[e.g.,][]{1983SoPh...87..243M} or semy manually operated \citep[e.g.,][]{2006SoPh..237...13M} algorithms are impractical for processing significantly large data sets.
Although a few automated identification algorithms exist \citep[e.g.,][]{1993ApJ...415..832S,2007SoPh..tmp..109B} more identification methods used in previous works were semi-automated \citep[see e.g.,
][]{2006SoPh..237...13M}, i.e., the features
must be identified manually and are then tracked automatically. This method is
quite time-consuming and can be subjective, and a more efficient and objective method of identification and feature tracking is therefore required. Different approaches
to automated algorithms exist. Some authors attempted applying
Fourier-filtering techniques prior to identifying MBPs by given thresholds. Others developed single level
(of brightness) identification algorithms. An automated algorithm is also required for analysing long time series such as those obtained from the Hinode mission and reduces the uncertainty. The uncertainty can be reduced in two ways. Firstly, the statistical uncertainty
decreases with the increasing size of data sets. Secondly, results derived by an automated algorithm do not depend on the user
but rather on a specific parameter set. A further advantage would be the reproducibility of the results.
Therefore, we developed an automated identification algorithm
similar to the so-called MLT 4 algorithm of
\citet{2007SoPh..tmp..109B}, which was originally developed for investigatng granulation patterns and later adapted to investigations of MBPs. Our algorithm is dedicated to MBP research. Our approach, the resulting algorithm,
and our results derived so far, are outlined in this paper.

This paper is the first in a series of articles dealing with Magnetic Bright Points in the solar photosphere. Here, we concentrate on the developed identification algorithm.

\section{Data}
We used two data sets of the SOT \citep[Solar Optical Telescope; for a description see e.g.,][]{2004SPIE.5487.1142I,2008SoPh..tmp...26S} an instrument onboard
the Hinode mission \citep[e.g., see][]{2007SoPh..243....3K} built in cooperation between NAOJ (National Astronomical Observatory Japan) and Lookhead Martin corporation. The SOT has a
50 cm primary mirror, which limits the spatial resolution by diffraction
to be about 0.2 arcsec.

For our analysis two data sets of G-band images (here the contrast of MBPs is increased, compared to white-light images) close to disc centre were used. Data set I had a field of view (FOV) of 55.8 arcsec by 111.6 arcsec with a spatial sampling of 0.108 arcsec/pixel. The complete time series consisted of 645
G-band images. Data set II had a FOV of 27.7~arcsec by
27.7~arcsec with a spatial sampling of 0.054 arcsec/pixel and consisted of 756 exposures.

Both data sets were fully
calibrated and reduced by Hinode standard data-reduction algorithms
distributed under Solar SoftWare (SSW)\footnote{The software package can be obtained via the webpage:  http://msslxr.mssl.ucl.ac.uk:8080/SolarB/AnalysisSoftware.jsp}.

\begin{figure}
    \centering
        \includegraphics[scale=0.58]{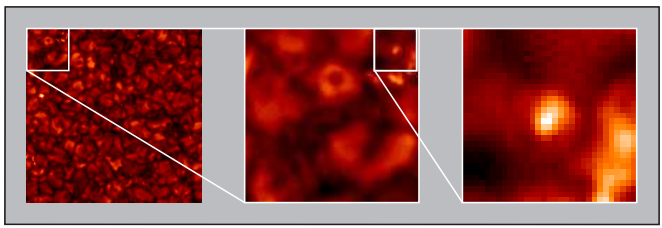}
        \caption{Left: Full-size Hinode/SOT exposure of data set II. The image has a size of about 27 by 27 arcsec.
        Middle: A zoomed part of this image is shown (size 130 by 130 pixels equal to be about 6 by 6 arcsec).
        Right: Another zoom in (31 by 31 pixels) which corresponds to be about 1.7 by 1.7 arcsec or about 1200 by 1200 km is displayed. One sees that the MBPs are very grainy (not smooth) structures.}
    \label{fig:detail2}
\end{figure}
\section{Algorithm}
The largest problem in identifying MBPs correctly with an automated algorithm is their small size (see the example Fig.~\ref{fig:detail2}) since some pixel noise always exists in the images that can be misinterpreted by an automated algorithm as small features. Furthermore, the small size has a significant influence on derived parameters such as areas or velocities.
Several characteristics of MBPs can be used by a
detection algorithm:
\begin{itemize}
    \item Brightness (MBPs are brighter than their immediate vicinity)
    \item Brightness gradient (MBPs should have a large brightness gradient)
    \item Size of features (MBPs are very small-scale features).
\end{itemize}
An identifying method on the basis of only one of those MBP characteristics would be unsuccessful for the following reasons. For
example, if the only criterion for a feature to be identified as MBP was either brightness gradient or brightness, small granules and granular brightenings
would also be classified as MBPs. MBPs have a steep brightness
gradient since they arise out of the intergranular dark lanes, although, granules may also have a steep
brightness gradient at the boundary and a nearly constant brightness plateau.

\begin{figure}
    \centering
    \includegraphics[scale=0.35]{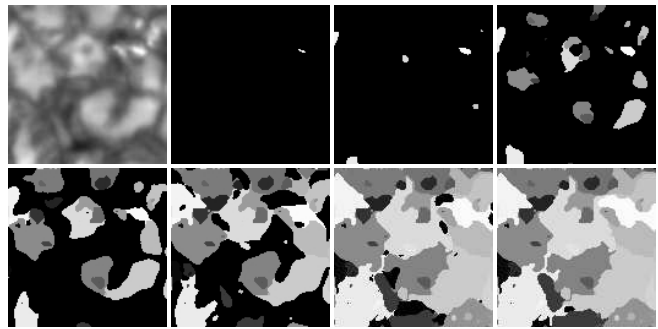}
        \caption{Illustration of the segmentation algorithm.
        From top left to bottom right, every 40th segmentation step is shown. The image used here for illustration is a subimage of one Hinode/SOT exposure, showing a field of view of 10.8 arcsec by 10.8 arcsec.}
    \label{fig:areas_image235}
\end{figure}

\begin{figure*}
    \centering
        \includegraphics[width=\textwidth]{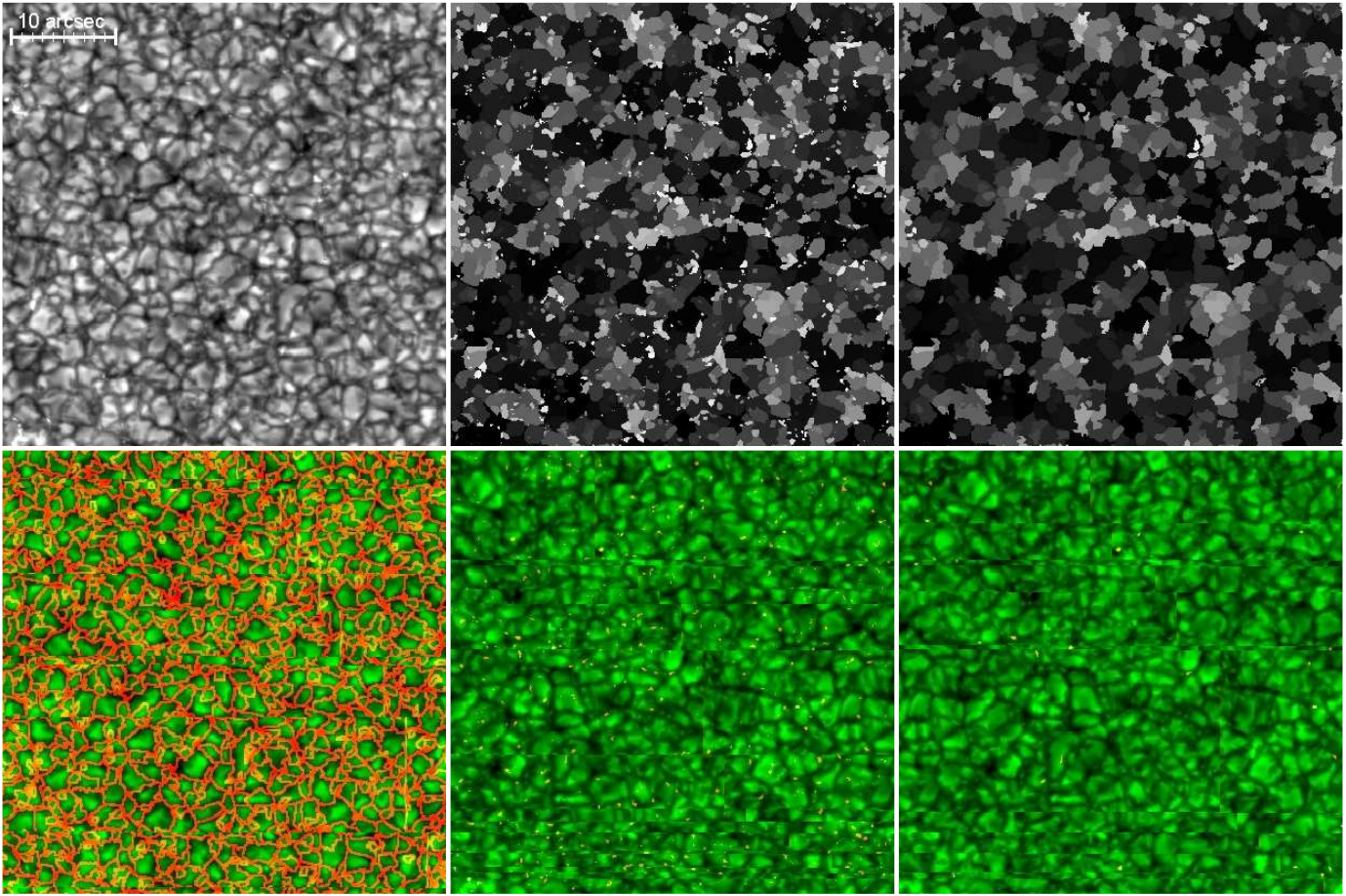}
        \caption{Left top figure shows the original image (spatial sampling of 0.108 arcsec/pixel). Several bright
        points could be seen. Next, the segmented image is shown before the application of the clean-up algorithm. Followed by the image of the segments after application of the clean-up algorithm. On the left bottom side, the false color image is plotted with the derived feature borders. The middle bottom panel shows the image and the pixels identified as Bright Points (without application of a clean up). Finally, the identified features after application of the clean-up algorithm are plotted.}
    \label{fig:testneu1}
\end{figure*}

Therefore, we have to derive the size of the features, which requires a segmentation algorithm. Since there will be some oversegmentation, the size of the features alone will be insufficient to ensure
that the identified feature is truly a MBP. We therefore require an algorithm for size detection
(segmentation), which also uses the information about the brightness and
brightness gradient.
The developed algorithm consists of three major processing steps:
\begin{itemize}
  \item Segmentation step
  \item Clean-up step
  \item Identification step
\end{itemize}
The segmentation algorithm is based on a region-growing algorithm,
i.e., on the idea of following the border of a feature from the
brightest pixels to the faintest. For this
purpose, the brightest pixel of an image is taken and initialises the
first feature. In the next step, pixels of slightly lower
brightness than in the first step are taken and merged into the first
feature if they are neighboring this feature. Otherwise, they
initialise the next feature. By repetition of these steps, the
image becomes segmented into single features. A pixel that has to be assigned, can often border two features, so the algorithm must decide to which of the features the pixel
should belong. In our case, the pixel is assigned to the feature created first, and oversegmented parts
of features then tend to stay small. However, brighter features may tend to spread out into intergranular regions. Figure \ref{fig:areas_image235}, illustrates the different steps of the segmentation routine, and Fig. \ref{fig:testneu1} shows the results after segmentation.

When the algorithm finishs, the original image has transformed into a
segmented image. Each of the features
can then be accessed and analysed. At this stage, the segmentation algorithm may also oversegment some features i.e., the segmentation process may have split single features if they possess several brightness maxima. 

The clean up algorithm is applied to reduce the oversegmentation of
single features. There are basically three processes
that produce oversegmentation:
\begin{itemize}
  \item Pixel noise
  \item Cosmic rays and particle hits
  \item Granular brightenings
\end{itemize}
Pixel noise is transformed into small brightness
fluctuations, which in our case cause additional features to be
created in the segmentation process. Cosmic-ray hits also cause oversegmentation and could be minimised by applying despiking algorithms to the data. Those algorithms, however, tend to remove small-scale features (such as our GBPs), and we therefore did not apply these algorithms. If granules have several brightness maxima (granular brightenings), the algorithm will split these granules into several segments.

The difference between the brightest pixel and the brightest border
pixel of a feature was calculated. If this value is smaller than a certain
threshold value, the feature is most probably an oversegmented
element on top of a granular feature (i.e., a granular brightening). However,
it could also be a bright point on the edge of a granule and a second difference (between brightest pixel and mean border
brightness) is therefore calculated. If the segmented feature is a MBP or a bright granular border, this difference should be large, since the mean
border brightness of the MBP should be small (most border pixels lie
in the dark intergranular region; only a few connect the MBP
to the granule). If this derived value is below a second
threshold, the feature will be merged with the element that has
the brightest bordering pixel. The result of this process can be seen in Fig.
\ref{fig:testneu1}.

\begin{figure}
    \centering
         \includegraphics[scale=0.3]{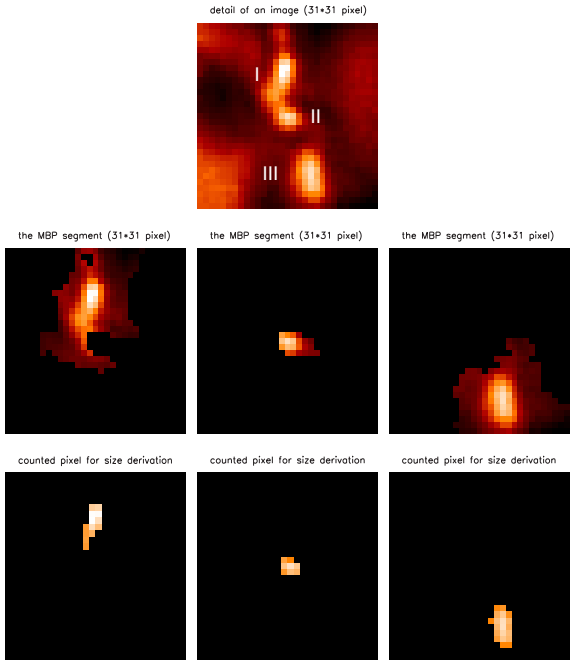}
        \caption{Demonstration of the size determination. Top: A 31 by 31 pixels zoom in detail of an image is shown. Second row: The segment of the image which belongs to the MBP feature (I left, II middle, III right) is displayed. If all of the segment pixels would be taken into account for the size of the MBP, the derived area would too large. Bottom row: The pixels actually taken into account for the size calculation.}
    \label{fig:area_compare}
\end{figure}

The final step (before analysis) is the identification of the MBPs, when the
size of each image segment was calculated. We must first define "size" appropriately. The
simplest idea would be to take all of the pixels of a feature
segment into account. This is problematic, since a feature quite
often extends continuously into the intergranular region, and
therefore the derived size would be far too large (a visualisation
of this can be seen in Fig. \ref{fig:area_compare}). In most studies, the size was defined by considering all pixels exceeding a certain brightness level. A disadvantage of this approach is missing fainter MBPs, i.e., all of the features that are bright relative to their vicinity but not sufficiently bright to exceed this
criterion. Our definition conflicts with these definitions. We consider all pixels in a segment that have brightnesses between an upper and lower boundary. The upper boundary is given by the brightest pixel in the segment. The lower boundary is given by the upper boundary minus a threshold parameter (set in the program). The brightest pixel in a segment has, for example, a value a factor of 1.5 above the mean photospheric intensity and the brightness threshold is set to be 30\%. All of
the pixels of brightness between a factor of 1.5 to 1.2 of the segment are taken into
account in calculating the size. MBPs can be described by
flux tubes. The inner part should be homogenously bright, and on the
outside the MBPs should be surrounded by dark intergranular matter.
This produces a significant brightness gradient in a small
transition region between the intergranular and MBP region. In our case,
this transition region is not taken into account for the size
determination. Because of the steep intensity gradient, slight
variations in the selected brightness threshold do not correspond to significant differences in the derived sizes (changing the parameter
by 30 percent changes the derived sizes by about 10
percent). Finally, the correct elements are selected on the basis of their sizes. Each feature that does not exceed a certain upper cut-off size is regarded to be a MBP. In our case, these values were 10 and 25 pixels, respectively (see also Fig. \ref{fig:size_distribution_2}). Information about these
features, such as brightness, size, position, and image number (equal
to time), are recorded separately. Finally, important parameters such as
the size distribution or the number density of MBPs in the Field of
View (FOV) are calculated.

\section{Results}
\begin{figure}
    \centering
        \includegraphics[scale=0.45]{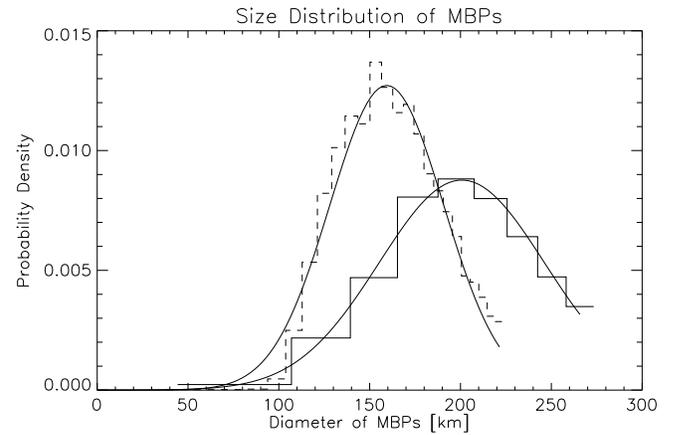}
        \caption{MBP size distributions for both data sets and their fitted normal distributions.
        Data set I displayed in solid line (spatial sampling of 0.108 arcsec/pixel). Data set II displayed in dashed line (sampling rate of 0.054 arcsec/pixel). Fit parameters of data set I are 218 $\pm$ 48 km and fit parameters for data set II are 166 $\pm$ 31 km.}
    \label{fig:size_distribution_2}
\end{figure}

The size distribution was determined by deriving the area of each
MBP in pixels. The size criterion was chosen in such a way that
only the brightest pixels of the MBPs were taken into account for
size derivation. The threshold was set to be 30\% of the mean
photospheric intensity. This value was chosen because the automated processed images had then the closest agreement when compared to visually identified and processed ones. The derived area was then transformed into a
circle of equal area.
The size distribution obtained from data set I
(sampling of 0.108$\arcsec$ per pixel), follows a normal
distribution with the fit parameters $\sigma~=~48~\mathrm{km}$
(standard deviation) and $\mu~=~218~\mathrm{km}$ (mean value). This size
distribution incorporates in total
52~628 measurements of MBP sizes in 645 G-band time-series images. The size distribution derived for data set II (sampling of 0.054$\arcsec$ per pixel), also follows a normal
distribution but with different fitting parameters of $\sigma~=~31~\mathrm{km}$ and $\mu~=~166~\mathrm{km}$. In this case,
7~891 single MBP size measurements were obtained for 756 G-band time-series images. The ratio of the
number of measurements in both data sets I and II (of about 7) corresponds to the difference in FOV
(27$\arcsec$ by 27$\arcsec$ to 55$\arcsec$ by 111$\arcsec$). Despite the lower number of detected MBPs, the data set of higher spatial resolution is more reliable because it is ``closer'' to the true size distribution due to its smaller pixel size. Both
distributions are shown in Fig. \ref{fig:size_distribution_2}. It is
obvious that the distribution is shifted to smaller values.

Since the algorithm has problems in differentiating between large MBPs and small granules, we could only derive sizes of up to 280 km and 220 km, respectively.

\begin{figure}
	\centering
\includegraphics[scale=0.45]{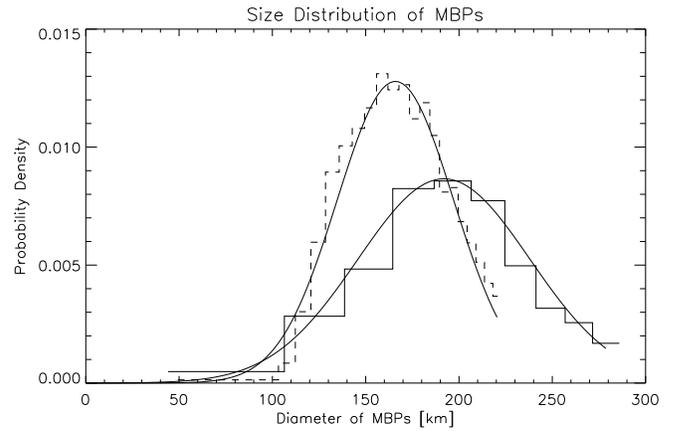}
	\caption{The size distribution of data set II together with the size distribution derived after decreasing the spatial sampling artificially. Original fitting parameters: 166 $\pm$ 31 km. Fitting parameters afterwards: 192 $\pm$ 46 km.}
	\label{fig:size_distribution_deg}
\end{figure}

The shift in the two size distributions is most likely caused by the change in the sampling resolution (changed from 0.108 arcsec/pixel to 0.054 arcsec/pixel). We checked this by artificially decreasing the sampling resolution of data set II and then applied our algorithm again using the same set of parameters. The result can be seen in Fig. \ref{fig:size_distribution_deg}. For the artificially lower resolved data set (which corresponds to a lower spatial sampling), we found values of about ($192\pm46$) km that agree with the above presented Fig. \ref{fig:size_distribution_2}. The degradation in sampling was achieved by summing the intensity values of four neighbouring pixels to derive that of the new single pixel.

We now examine the effects of discretisation. The size distribution found for data set II (166 $\pm$ 31 km) is similar to that of
\citet{2004A&A...422L..63W}. Their work was based on data from the 1
m Swedish Solar Telescope (SST) on La Palma with a sampling of 0.041$\arcsec$ per pixel. They derived most frequently a diameter of 160 $\pm$ 20 km. We consider the shift and change in our size distribution with sampling to be due to an undersampling of the features. This means that the algorithm derives sizes that are larger than in reality. 
The sizes could not only be overestimated but also
underestimated, if a feature is only recognised partly. This is due to an averaging effect during the discretisation. Every pixel value represents the average brightness of the corresponding true surface area on the sun. For small features (occupying a few pixels), this effect can lead to the feature being completely missed, particularly when the MBPs are considered to be located in the dark intergranular region. Therefore, small bright elements would be averaged with dark intergranular areas, and the features would not longer be identified as a bright element. This process produces the shift in the distribution to larger sizes in the case of lower sampling (Fig. \ref{fig:size_distribution_2}). If there is a certain probability of deriving values that are too large or too small, a not so well sampled distribution will become broader. This is because more values will be redistributed from the center into the wings of the distribution than vice versa. We therefore conclude that the sampling has a high influence
on the observed parameters of small-scale features, which we already demonstrated above (Figs. \ref{fig:size_distribution_2} and \ref{fig:size_distribution_deg}). We conclude that the size
distribution will converge to the true distribution, when not only the true diffraction limit improves but also the spatial sampling is increased.

Sizes around and below the theoretical resolution limit (in our case of between 140 km and 160 km) have to be doubted because these regions could normally never be resolved. Sizes found below the theoretical resolution limit are most probably due to uncertainties (e.g. pixel noise, and algorithm artefacts).
\begin{figure}
    \centering
        \includegraphics[scale=0.40]{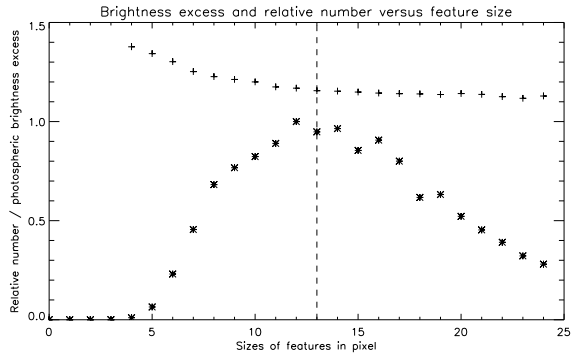}
        \caption{The relative number and brightness excess (to the mean photospheric brightness) versus the size of the
        features in pixel is shown. Only values right of the vertical line at 13 pixels (diffraction limit) should be relied on.}
    \label{fig:bright_size2}
\end{figure}
\begin{figure}
    \centering
        \includegraphics[scale=0.40]{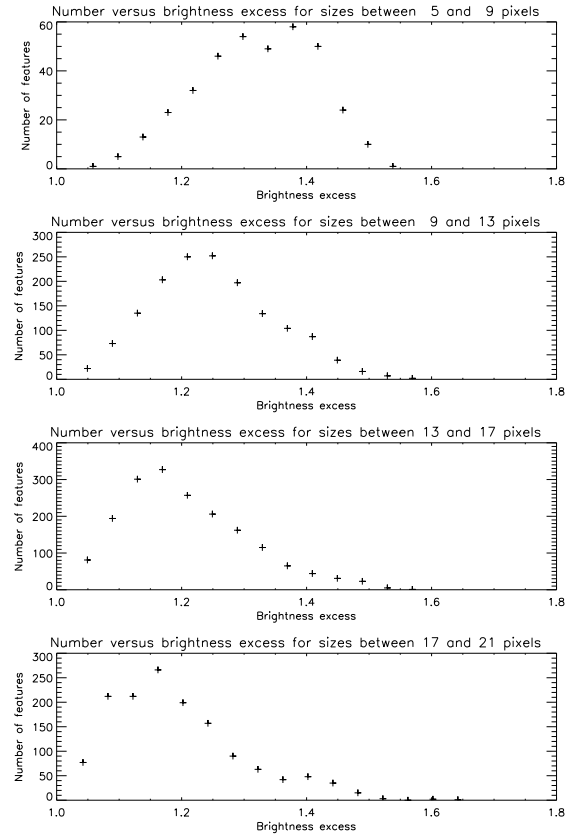}
        \caption{The number versus brightness excess distribution shifts to higher brightness excesses, when the detected feature sizes decrease.}
    \label{fig:bright_number}
\end{figure}
To deal with oversegmentation, the detected elements must have at least a defined minimum difference between their brightest
 pixels and their mean border pixels. When sizes become smaller and smaller, we reach a regime where the brightest pixel is also a border pixel of the feature (below 9 pixels this is always true). If the pixel brightness increases, the mean border brightness of the entire feature increases. Therefore, the brightest pixel in the feature must be considerably brighter to maintain the brightness threshold criteria. This
  behavior is illustrated in Figs. \ref{fig:bright_size2} and \ref{fig:bright_number}. We conclude that this effect steepens the
  decrease in the size distribution from the maximum to smaller sizes. The left branch of the distribution is in all the cases below the theoretical resolution limit of the telescope. Our investigation suggests that there is most probably no (or at least no
strong) brightness/size relation at all (in the accessible range down to about 140 km), as stated before by
\citet{1995ApJ...454..531B}.

\section{Conclusion}
In this article, we have described our new pattern recognition and
automated identification algorithm for MBPs in the solar
photosphere. We have demonstrated the potential of our algorithm and showed first results of the algorithm applied to data obtained from Hinode/SOT. The derived size
distribution of the MBPs depends strongly on the achieved
resolution. Therefore, we suppose that the spatial resolution of Hinode/SOT is insufficient to resolve the smallest elements in the photosphere and that higher resolution images
are needed such as those obtained by the 1m Swedish Solar Telescope (SST) on La Palma/Spain. Nevertheless the results agree well
with earlier published values and we are confident that our algorithm is suitable for analysing not only static parameters
such as size distributions but also dynamical characteristics such as velocities and
lifetimes. Insight into the sampling problem could be gained
by computer simulations. In the future, it will be an important task to derive size distributions with computer-simulation algorithms
and compare them with observational data to derive a more robust understanding of errors arising in the sampling
processes.

We emphasise that the influence of discretisation is of course not only unique to MBPs but a general problem for other investigations (dealing with similar sampled data) as well. Therefore, this effect should be considered and investigated in more detail.

\acknowledgements
We are grateful to the Hinode team for the possibility to use their data.
Hinode is a Japanese mission developed and launched by ISAS/JAXA, with NAOJ as domestic partner and NASA and STFC (UK) as international partners. It is operated by these agencies in co-operation with ESA and NSC (Norway).
This work was supported by FWF \emph{Fonds zur F{\"o}rderung wissenschaftlicher Forschung} grant: P20762
D. U. and A. H. are grateful to the {\"O}AD \emph{{\"O}sterreichischer Austauschdienst} for financing a scientific stay at the Pic du Midi Observatory. R.M. is grateful to the Minist\`{e}re des Affaires Etrang\`{e}res et Europ\'{e}ennes,
for financing a stay at the University of Graz.
We want to thank the anonymous referee for his remarks and comments, which helped us to improve the outcome of this work.

\bibliographystyle{aa}

\begin{thebibliography}{9}
\expandafter\ifx\csname natexlab\endcsname\relax\def\natexlab#1{#1}\fi

\bibitem[{{Berger} {et~al.}(1995){Berger}, {Schrijver}, {Shine}, {Tarbell},
  {Title}, \& {Scharmer}}]{1995ApJ...454..531B}
{Berger}, T.~E., {Schrijver}, C.~J., {Shine}, R.~A., {et~al.} 1995, \apj, 454,
  531

\bibitem[{{Bovelet} \& {Wiehr}(2007)}]{2007SoPh..tmp..109B}
{Bovelet}, B. \& {Wiehr}, E. 2007, \solphys, 109

\bibitem[{{Ichimoto} {et~al.}(2004){Ichimoto}, {Tsuneta}, {Suematsu},
  {Shimizu}, {Otsubo}, {Kato}, {Noguchi}, {Nakagiri}, {Tamura}, {Katsukawa},
  {Kubo}, {Sakamoto}, {Hara}, {Minesugi}, {Ohnishi}, {Saito}, {Kawaguchi},
  {Matsushita}, {Nakaoji}, {Nagae}, {Sakamoto}, {Hasuyama}, {Mikami},
  {Miyawaki}, {Sakurai}, {Kaido}, {Horiuchi}, {Shimada}, {Inoue}, {Mitsutake},
  {Yoshida}, {Takahara}, {Takeyama}, {Suzuki}, \& {Abe}}]{2004SPIE.5487.1142I}
{Ichimoto}, K., {Tsuneta}, S., {Suematsu}, Y., {et~al.} 2004, in Presented at
  the Society of Photo-Optical Instrumentation Engineers (SPIE) Conference,
  Vol. 5487, Optical, Infrared, and Millimeter Space Telescopes. Edited by
  Mather, John C. Proceedings of the SPIE, Volume 5487, pp. 1142-1151 (2004).,
  ed. J.~C. {Mather}, 1142--1151

\bibitem[{{Kosugi} {et~al.}(2007){Kosugi}, {Matsuzaki}, {Sakao}, {Shimizu},
  {Sone}, {Tachikawa}, {Hashimoto}, {Minesugi}, {Ohnishi}, {Yamada}, {Tsuneta},
  {Hara}, {Ichimoto}, {Suematsu}, {Shimojo}, {Watanabe}, {Shimada}, {Davis},
  {Hill}, {Owens}, {Title}, {Culhane}, {Harra}, {Doschek}, \&
  {Golub}}]{2007SoPh..243....3K}
{Kosugi}, T., {Matsuzaki}, K., {Sakao}, T., {et~al.} 2007, \solphys, 243, 3

\bibitem[{{M{\"o}stl} {et~al.}(2006){M{\"o}stl}, {Hanslmeier}, {Sobotka},
  {Puschmann}, \& {Muthsam}}]{2006SoPh..237...13M}
{M{\"o}stl}, C., {Hanslmeier}, A., {Sobotka}, M., {Puschmann}, K., \&
  {Muthsam}, H.~J. 2006, \solphys, 237, 13

\bibitem[{{Muller} \& {Keil}(1983)}]{1983SoPh...87..243M}
{Muller}, R. \& {Keil}, S.~L. 1983, \solphys, 87, 243

\bibitem[{{Sobotka} {et~al.}(1993){Sobotka}, {Bonet}, \&
  {Vazquez}}]{1993ApJ...415..832S}
{Sobotka}, M., {Bonet}, J.~A., \& {Vazquez}, M. 1993, \apj, 415, 832

\bibitem[{{Suematsu} {et~al.}(2008){Suematsu}, {Tsuneta}, {Ichimoto},
  {Shimizu}, {Otsubo}, {Katsukawa}, {Nakagiri}, {Noguchi}, {Tamura}, {Kato},
  {Hara}, {Kubo}, {Mikami}, {Saito}, {Matsushita}, {Kawaguchi}, {Nakaoji},
  {Nagae}, {Shimada}, {Takeyama}, \& {Yamamuro}}]{2008SoPh..tmp...26S}
{Suematsu}, Y., {Tsuneta}, S., {Ichimoto}, K., {et~al.} 2008, \solphys, 26

\bibitem[{{Wiehr} {et~al.}(2004){Wiehr}, {Bovelet}, \&
  {Hirzberger}}]{2004A&A...422L..63W}
{Wiehr}, E., {Bovelet}, B., \& {Hirzberger}, J. 2004, \aap, 422, L63

\end{thebibliography}

\end{document}